\documentclass[aps,twocolumn,superscriptaddress,longbibliography]{revtex4-1}

\usepackage{graphicx,xcolor}
\usepackage{amssymb,bbm}
\usepackage[reqno]{amsmath}
\usepackage{footmisc}
\usepackage[colorlinks,bookmarks=false,citecolor=blue,linkcolor=red,urlcolor=blue,hyperfootnotes=true]{hyperref}

\begin{document}

\title{Non-local Coulomb correlations in pure and electron-doped
  ${\mathrm{Sr}}_{2}{\mathrm{IrO}}_{4}$:
  spectral functions, Fermi surface and pseudogap-like spectral weight
  distributions from oriented cluster dynamical mean field theory
}

\author{Cyril Martins}
\thanks{These two authors contributed equally to this work.}
\affiliation{Laboratoire de Chimie et Physique Quantiques, UMR 5626, Universit{\'{e}} Paul Sabatier, 118 route de Narbonne, 31400 Toulouse, France}
\author{Benjamin Lenz}
\thanks{These two authors contributed equally to this work.}
 \affiliation{Centre de Physique Th{\'{e}}orique, Ecole Polytechnique, CNRS UMR 7644, Universit{\'{e}} Paris-Saclay, 91128 Palaiseau, France}
\author{Luca Perfetti}
\affiliation{Laboratoire des Solides Irradi\'es, Ecole Polytechnique, CNRS, CEA, Universit{\'{e}} Paris-Saclay, 91128 Palaiseau, France}
\author{Veronique Brouet}
\affiliation{Laboratoire de Physique des Solides, Universit\'e Paris-Sud, Universit\'e Paris-Saclay, 91405 Orsay, France}
\author{Fran\c{c}ois Bertran}
\affiliation{Synchrotron SOLEIL, L'Orme des Merisiers, Saint-Aubin-BP 48, 91192 Gif sur Yvette, France}
\author{Silke Biermann}
 \email{silke.biermann@polytechnique.edu}
 \affiliation{Centre de Physique Th{\'{e}}orique, Ecole Polytechnique, CNRS UMR 7644, Universit{\'{e}} Paris-Saclay, 91128 Palaiseau, France}
\affiliation{Coll{\`{e}}ge de France, 11 place Marcelin Berthelot, 75005 Paris, France}

\date{\today}                 

%
\newcommand{\eref}[1]{equation~(\ref{#1})}
\newcommand{\fref}[1]{figure~\ref{#1}}
\newcommand{\Fref}[1]{Figure~\ref{#1}}
\newcommand{\tref}[1]{table~\ref{#1}}
\newcommand{\green}[1]{\textcolor{green}{#1}}
\newcommand{\blue}[1]{\textcolor{blue}{#1}}
\newcommand{\im}{%
           \imath}
\newcommand{\bra}[1]{\ensuremath{\langle #1|}}
\newcommand{\ket}[1]{\ensuremath{|#1\rangle}}
\newcommand{\braket}[2]{\langle #1|#2\rangle}
\newcommand{\bracket}[2]{\langle #1|#2\rangle}
\newcommand{\bbra}[1]{\ensuremath{\big\langle #1\bigr|}}
\newcommand{\bket}[1]{\ensuremath{\bigl|#1\big\rangle}}
\newcommand{\Bbra}[1]{\ensuremath{\bigg\langle #1\biggr|}}
\newcommand{\Bket}[1]{\ensuremath{\biggl|#1\bigg\rangle}}
\newcommand{\up}{%
        \ensuremath{\uparrow}}
\newcommand{\down}{%
        \ensuremath{\downarrow}}
\newcommand{\updown}{%
        \ensuremath{\uparrow\!\downarrow}}
\newcommand{\fermi}[1]{%
        \hbox{f($#1$)}}
\newcommand{\dif}{%
        \hbox{d}}
\newcommand{\tr}{%
        \hbox{ tr}}
\def\beq{\begin{equation}}
\def\eeq{\end{equation}}
\def\vk{{\bf k}}
\def\eg{{e.g.}}
\def\egp{\ensuremath{e_g^\pi}}
\def\egs{\ensuremath{e_g^\sigma}}
\def\t2g{\ensuremath{t_{2g}}}
\def\a1g{\ensuremath{a_{1g}}}
\def\se{self-energy}
\newcommand{\red}[1]{\textcolor[rgb]{0.71, 0.07, 0.26}{#1}}
\newcommand{\cc}{%
        c^\dag}
\newcommand{\ca}{%
        c^{\phantom{\dag}}}  
\newcommand{\svek}{%
        \mathbf}
\newcommand{\vek}[1]{%
        \hbox{\textbf #1}}
\newcommand{\op}[1]{%
        \hbox{\textbf #1}}
\newcommand{\pr}{%
        ^\prime}
\newcommand{\ppp}{%
        ...\\}
\newcommand{\K}{~\mbox{K}}
\def\vo2{VO\ensuremath{_2}}
\def\tio2{TiO\ensuremath{_2}}
\def\sio2{SiO\ensuremath{_2}}
\def\VO2{VO\ensuremath{_2}}
\def\v2o3{V$_2$O$_3$}
\def\V2O3{V$_2$O$_3$}
\def\vcro2{V$_{1-x}$Cr$_x$O$_2$}
\def\etal{{\it et~al.}}
\def\dparr{$d_{\parallel}$}
\def\hlda{H^{\rm{LDA}}}
\def\vk{{\bf k}}
\def\hn{\hat{n}}
\def\cc{\hat{c}^{\dagger}}
\def\ca{\hat{c}}
\def \bairo{Ba${}_2$IrO$_{4}$}
\def \sriro{Sr${}_2$IrO$_{4}$}
\def \lacuo{La${}_2$CuO$_{4}$}
\def \srruo{Sr${}_2$RuO$_{4}$}
\def \t2g{t$_{2g}$}
\def \eg{e$_{g}$}
\def \dxy{d$_{xy}$}
\def \dxz{d$_{xz}$}
\def \dyz{d$_{yz}$}
\def \dx2y2{d$_{x^2-y^2}$}
\def \j32{$j_{\textrm{eff}}$=$3/2$}
\def \jeff12{$j_{\textrm{eff}}$=$1/2$}

\def\delom{\delta\omega_{1/2}}
\def\delomc{\delta\omega_c}

\def\ve{\varepsilon}
\def\evk{\varepsilon_{v{\bf k}}}
\def\eck{\varepsilon_{c{\bf k}}}
\def\wvk{w_{v{\bf k}}}
\def\wck{w_{c{\bf k}}}


\begin{abstract}
We address the role of non-local Coulomb correlations and short-range
  magnetic fluctuations in the high-temperature phase of
Sr$_2$IrO$_4$ within state-of-the-art spectroscopic and 
first-principles theoretical methods.
Introducing a novel cluster dynamical mean field scheme,
we compute momentum-resolved spectral functions, which we find to be in
excellent agreement with angle-resolved photoemission spectra.
We show that while short-range antiferromagnetic fluctuations are crucial
to account for the electronic properties of the material even in the 
high-temperature paramagnetic phase, long-range magnetic order is not a
necessary ingredient of the insulating state.
Upon doping, an exotic metallic state is generated, exhibiting cuprate-like pseudo-gap spectral properties, for which we propose a surprisingly
simple theoretical mechanism.
\end{abstract}

\maketitle


\section{Introduction}
\label{Intro}

The iridium oxide Sr$_2$IrO$_4$ is not only isostructural to the celebrated
high-temperature superconducting copper oxides of the La$_2$CuO$_4$ family.
The similarities in the low-energy electronic structure of these compounds,
with a single orbital forming the low-energy states, are even more intriguing.
In the case of cuprates it is a single half-filled x$^2$-y$^2$ orbital, 
subject to strong electronic Coulomb correlations, that determines the 
low-energy properties, presumably including superconductivity \cite{KKN+15}.
In the t$_{2g}^5$ system Sr$_2$IrO$_4$ a complex spin-orbit entangled
compound orbital carries a single hole \cite{KJM+08,KOK+09}.
While in the cuprates the single-orbital nature of the low-energy
electronic structure results from the single-particle band structure,
in the iridate it is the result of the joined action of Coulomb interactions
and spin-orbit interactions that effectively suppress the degeneracy
\cite{MAVB11, MAB17}.

As in the cuprates, stoechiometric samples of Sr$_2$IrO$_4$
are insulating as a result of Coulomb interaction effects beyond the single-particle band picture, and at low temperatures antiferromagnetic order sets in \cite{CBM+98}. The phase diagram has been explored extensively
both, using different theoretical methods
\cite{MAVB11,JJOY09,JK09, WSY10, MAVB11,WS11, KKM12,AKK+12, ZHV13, MKK14, IN14, WSY14, LYJ+15,  SMK15, MAB17}
and
by various experimental probes:
optics \cite{MJK+08,KJM+08,MJC+09,HMT+12,ZTC+16,PYH+16,LSZ+16},
photoemission \cite{KJM+08,WCW+13,KKD+14,YTF+14,BMP+15,dlTMWB+15,LYJ+15,NKK+15,KUP+16,PBP+16,CWW+16,KSDK16}, 
transport \cite{KOK+09, CKC+09, GQK+11, ZHW+14, FCCB15, PBB+16, ZGY+17},
scanning tunneling microscopy and spectroscopy experiments  \cite{LCO+13, DCCM14}, and X-ray spectroscopy \cite{HFZ+12,FOK+12,KCU+12,FOO+14, KDS+14}
have contributed to establish
the picture of an insulator with a strongly
temperature-dependent gap that is however not affected by the onset of
magnetic order. 
This is {\it a fortiori} intriguing since even in the paramagnetic phase long-range two-dimensional antiferromagnetic fluctuations exist, which have been found to exceed 100 lattice spacings \cite{FOK+12}.
Magnetic exchange interactions seem to 
be of correspondingly long-range nature \cite{AKMS+17}.
First principles dynamical mean field theory (DMFT) calculations \cite{MAVB11}
can rationalize the insulating nature of the compound, even in its
paramagnetic phase. This is consistent with the experimentally found
insensitivity of spectral or transport properties on the presence of
absence of magnetic order \cite{QKL+12,BMP+15,ZTZ+17}.
Nevertheless the detailed analysis of band dispersions obtained within
single-site DMFT  
reveals interesting discrepancies between theory and experiment (see below).
For the case of the related Ba$_2$IrO$_4$ \cite{HPL15} 
cluster DMFT leads to much better agreement with angle-resolved 
photoemission data \cite{MME+14,UNK+14}.

Following the analogy with the cuprates, one of the most intriguing
questions is the evolution of the electronic structure when
additional carriers are introduced.
Recently, several groups have succeeded to dope Sr$_2$IrO$_4$,
both on the electron- and hole-doped side. Most interestingly, however,
despite all the analogies with the cuprates, to date no superconducting
phase could be observed.

A metallic state can be realized via cationic substitution both in the case of hole doping, e.g. with K \cite{GQK+11}, Rh \cite{KT08,QKL+12,LKT+12} or Ru \cite{CKC+15}, and in the case of electron doping with La \cite{KT08,GQK+11,CHW+15,ZTZ+17,BFB+17}.
For La doping $x\geq0.04$ the antiferromagnetic N\'eel temperature vanishes and the system $({\mathrm{Sr}}_{1-x}{\mathrm{La}}_x)_{2}{\mathrm{IrO}}_{4}$ remains a paramagnetic metal down to lowest temperatures \cite{GQK+11}.
Other possibilities of electron doping are depositing a K surface layer \cite{KKD+14,KSDK16} and via oxygen depletion \cite{KQC+10}.

Angle-resolved photoemission spectroscopy (ARPES) has been able to identify
the band dispersions and Fermi surfaces in electron- \cite{BMP+15,dlTMWB+15,PBP+16,KSDK16}
and hole-doped \cite{BMP+15,CWW+16,PBP+16} samples.
On the electron-doped side, the dominant Fermi surface feature is a
lens-shaped electron pocket centred around the $M$-point of the
crystallographic Brillouin zone. A detailed analysis of the onset of
spectral weight suggests the existence of depletion regions \cite{BMP+15,dlTMWB+15}
reminiscent of the pseudogap behavior observed in the cuprates \cite{AOM89,DYC+96,LSD+96}.

In this paper, we establish a first principles description of the
spectral properties of pure and electron-doped Sr$_2$IrO$_4$ beyond
the DMFT approximation of a purely local
many-body self-energy. We demonstrate that including short-range
fluctuations is crucial to reliably assess spectral properties,
which we find in excellent agreement with experiments.
In the electron-doped case we find an exotic metallic state, whose properties
we relate to recent angle-resolved photoemission spectra. 
In particular, our calculations offer a surprisingly simple picture
for the putative antinodal pseudogap found in experiments.
It is in fact a direct consequence of strong inter-site Ir-Ir
fluctuations.
Our findings suggest that while the similarities in the electronic
structure between iridates and cuprates cover various quite different
aspects, these common features may not be considered as proxy for
superconductivity.

\section{Assessing electronic properties of Sr$_2$IrO$_4$}
\label{ModTec}

\begin{figure*}[ht!]
\includegraphics[width=1.0\textwidth]{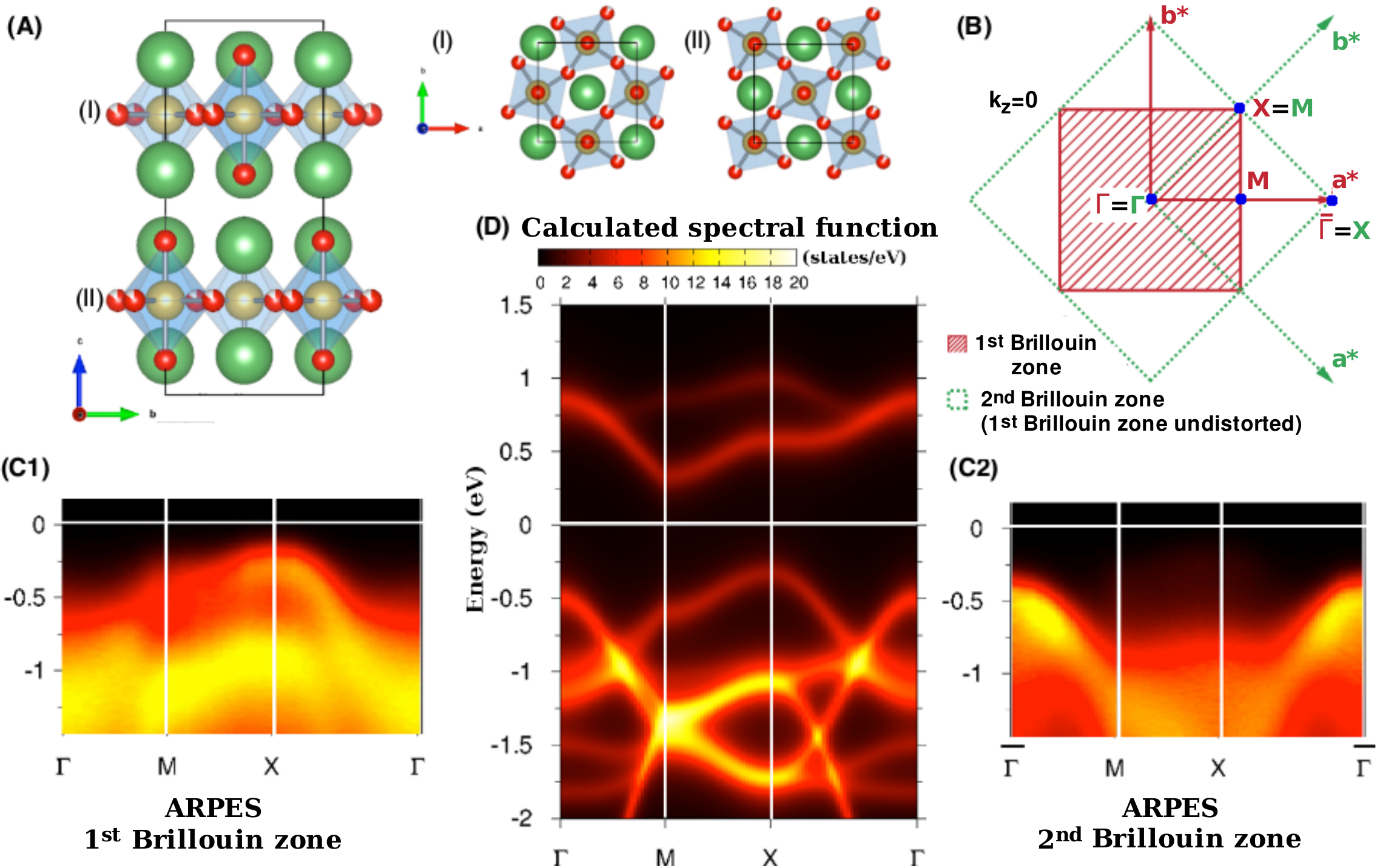}
\caption{\label{Akw}Spectral density of undoped ${\mathrm{Sr}}_{2}{\mathrm{IrO}}_{4}$. 
A: Crystal structure of $\mathrm{Sr}_2\mathrm{IrO}_4$ where green spheres stand for strontium ions, golden ones for iridium and blue ones for oxygen. The IrO$_6$ octahedra are alternately tilted clockwise and anticlockwise, leading to two different
-- though crystallographically equivalent -- configurations of the $\mathrm{Ir}$ atoms. The unit cell comprises two layers in the $c$-direction, which are shifted by $(1/2,0)$ in the $a-b$ plane. 
B: Illustration of the first Brillouin zone of the distorted and undistorted structure of $\mathrm{Sr}_2\mathrm{IrO}_4$. Rotations of the oxygen octahedra cause a doubling of the unit cell, which leads to a halved first Brillouin zone and a redefinition of high symmetry points. The high symmetry points of the distorted structure are shown in red. 
C: Measured ARPES spectrum in the first (C1) and second (C2) Brillouin zone. 
D: Calculated momentum-resolved spectral function of $\mathrm{Sr}_2\mathrm{IrO}_4$ using DFT and oriented-cluster DMFT for the $j_{\mathrm{eff}}=1/2$ band.}
\end{figure*}

The 5d transition metal oxide Sr$_2$IrO$_4$ crystallizes
in a tetragonal crystal structure 
derived from the K$_2$NiF$_4$ structure,
well-known in \srruo\ or \lacuo,  by lowering the
symmetry by a $11^\circ$ rotation of its IrO${}_{6}$ octahedra
around the $c$-axis \cite{HSC+94,MME+14}\footnote{Further reduction of the symmetry by staggered tetragonal distortion of the IrO${}_{6}$ octahedra \cite{TCZ+15} has not been considered here.}.
Below $240$~K, canted antiferromagnetic (AF) order sets in\cite{CBK+94,CSH+94,CBM+98}.
Here, we focus on the paramagnetic (PM) insulating phase of Sr$_2$IrO$_4$ above $240$~K.
The rotations of the IrO${}_{6}$ octahedra result in a doubling of the
unit cell along the $c$-direction, and a primitive cell in the paramagnetic
phase that is as large as in the antiferromagnetic one.
Its four crystallographically equivalent Ir atoms are pairwise related by
a screw-symmetry, with the orientation of pairs being rotated by 90 degrees
between neighboring planes, see Fig.~\ref{Akw}A and B.

DMFT-based first principles calculations
which introduce effective Hubbard interactions and assess the
resulting quantum fluctuations locally on each Ir atom could
indeed identify the insulating state of Sr$_2$IrO$_4$ \cite{MAVB11, MAB17},
even in the absence of long-range magnetic order.
The resulting description of spectral properties (see Fig.~3c of
Ref.~\cite{MAVB11}, which is replotted as Fig.~3
in the Supplemental Material\footnote{ See Supplemental Material at [URL will be inserted by publisher] }) is however not in agreement with experiment.

Here, we introduce
an extension of {\it ab initio}
cluster DMFT that extends the concept to
{\it oriented clusters} as
representative entities of parts of the solid. As in dynamical
mean field theory, a self-consistency condition restores the original
symmetries of the lattice, which -- in the case of an oriented cluster
-- concern however not only the usual translational symmetries, but
also the point group of the solid. Details of the method and its
practical implementation are discussed in the Supplemental Material \footnotemark[\value{footnote}].

Our starting Hamiltonian is the three-orbital Hubbard-type
Hamiltonian of Ref.~\cite{MAVB11}, where a
single-particle Hamiltonian for the t$_{2g}$-manifold is derived from
Density Functional Theory (DFT) calculations and augmented by Hubbard
and Hund's interaction terms.
As shown in Ref.~\cite{MAVB11}, however, the effect of the interactions onto
the \j32 states is essentially a global shift that makes these orbitals
completely filled, leaving only the \jeff12 states around the Fermi level.
Therefore, we take the result for the \j32 states directly from
Ref.~\cite{MAVB11}, and include only the \jeff12 states into our new
cluster theory treatment. This is done via a tight-binding parametrization
of the corresponding bands.

The oriented cluster DMFT (OC-DMFT) treatment of this Hamiltonian focuses onto the local Green's function
\begin{eqnarray}
  G_{loc} (\omega) = \sum_{\mathbf{k},\alpha} G_{\alpha}(\mathbf{k}, \omega)
\end{eqnarray}
with the momentum $\mathbf{k}$- and orientation $\alpha$-resolved Green's function 
\begin{eqnarray}
  G_{\alpha}(\mathbf{k}, \omega) = [\omega + \mu - H(\mathbf{k}) - \Sigma^{\alpha}_{dimer}(\omega)]^{-1}.
  \end{eqnarray}
Here, $\Sigma^{\alpha}_{dimer}$ is the self-energy of a dimer impurity
problem, augmented by an orientation $\alpha=a\pm b$, where $a$ and $b$ are the unit-cell vectors.
The $\mathbf{k}$ sum runs over the first Brillouin zone, restoring the
translational invariance of the solid after calculating its self-energy
from the quantum dimer problem, and the sum over $\alpha$ restores its
point group. More details on the general philosophy and practical
implementation can be found in the Supplemental Material \footnotemark[\value{footnote}].

In a metal, electronic screening is drastically enhanced as compared to an
insulator with profound consequences for the spectra \cite{ABW13,ABWB17}.
This is even more true in 5d compounds, where the relatively extended
nature of the 5d orbitals induces inter-atomic interactions that are relatively
large as compared to the local ones \footnote{H. Jiang et al., in preparation}.
For this reason, the effective Hubbard interactions are expected to be
smaller in the La-doped compound than in the pure sample.
We mimic this effect here by using a smaller on-site Hubbard interaction
for the doped case ($U_{\mathrm{eff}}=0.6$~eV) than for the undoped one ($U_{\mathrm{eff}}=1.1$~eV).

\section{Results}
\label{Result}

\subsection{Undoped Sr$_2$IrO$_4$}
\label{ResUndoped}

We have measured ARPES spectra of Sr$_2$IrO$_4$,
under the experimental conditions described in the Supplemental Material [67].
Figure~\ref{Akw}C displays the resulting spectra along the $\Gamma-M-X-\Gamma$
path in the first and second Brillouin zone. Due to matrix element effects,
these results display characteristic differences, with spectra in the first
Brillouin zone amplifying the \jeff12 contribution, while in the second
Brillouin zone the \j32 contribution is dominant. 
This strong matrix element effect expresses the fact that the corresponding Fourier
component of the potential (lowering the symmetry from $I_4 / mmm$ to $I_4 / acd$) is weak.
In agreement with ARPES spectra in the literature \cite{BMP+15,dlTMWB+15}, we find the
first removal state at $\Gamma$ to be of \j32 character, while the
\jeff12 form a strongly dispersive feature that displays a maximum at the
$X$ point.
In our theoretical calculations, we do not address matrix element effects,
which would differentiate results between different Brillouin zones.
For this reason, the theoretical spectral function is compared to the
sum of the experimental spectra.
Most intriguingly, in single-site DMFT calculations (see Fig.~3 in the Supplemental Material [67]), 
the \jeff12 states
form a very weakly dispersive feature with an onset of spectral weight at $-0.2$~eV below the Fermi level, in strong disagreement with the experimental spectra in Fig.~\ref{Akw}C.

Figure~\ref{Akw}D displays the result of our present calculations that
include non-local many-body correlations within our new OC-DMFT scheme.
The comparison to the experimental spectra 
yields impressive agreement, demonstrating that non-local many-body
effects were indeed the missing ingredient for assessing spectral
properties of this compound. The OC-DMFT treatment effectively
includes inter-iridium site fluctuations in the half-filled \jeff12
manifold, and in particular includes the inter-site magnetic exchange
of energy scale $4t^2/U$ into the description. With the present
parameters we obtain a value of $\sim108$~meV $(78$~meV) for states of inter-layer (anti)bonding nature, 
which coincides with the experimental estimate of the magnetic exchange coupling $J=4t^{2}/U\sim100$~meV \cite{FOK+12}.
As a consequence of the antiferromagnetic fluctuations, the \jeff12 band is much more dispersive than the single-site DMFT calculation and each of the Hubbard bands \footnote{From the dimer cluster used within our OC-DMFT treatment four spectral branches emerge: A bonding and antibonding band and two corresponding satellites, see also Ref.~\cite{Tom07}. Therefore one should rather refer to the branches with large spectral weight visible in the spectral plot of Fig.~\ref{Akw}D as (anti)bonding bands. Actually the correlation satellites or Hubbard bands are the outer excitations and carry less spectral weight. However, to be consistent with the notation in standard literature on Sr$_2$IrO$_4$ we use the notion of Hubbard bands when discussing the undoped compound.} has a width of $\sim0.8$~eV.

\begin{figure}[t]
\includegraphics[width=0.5\textwidth]{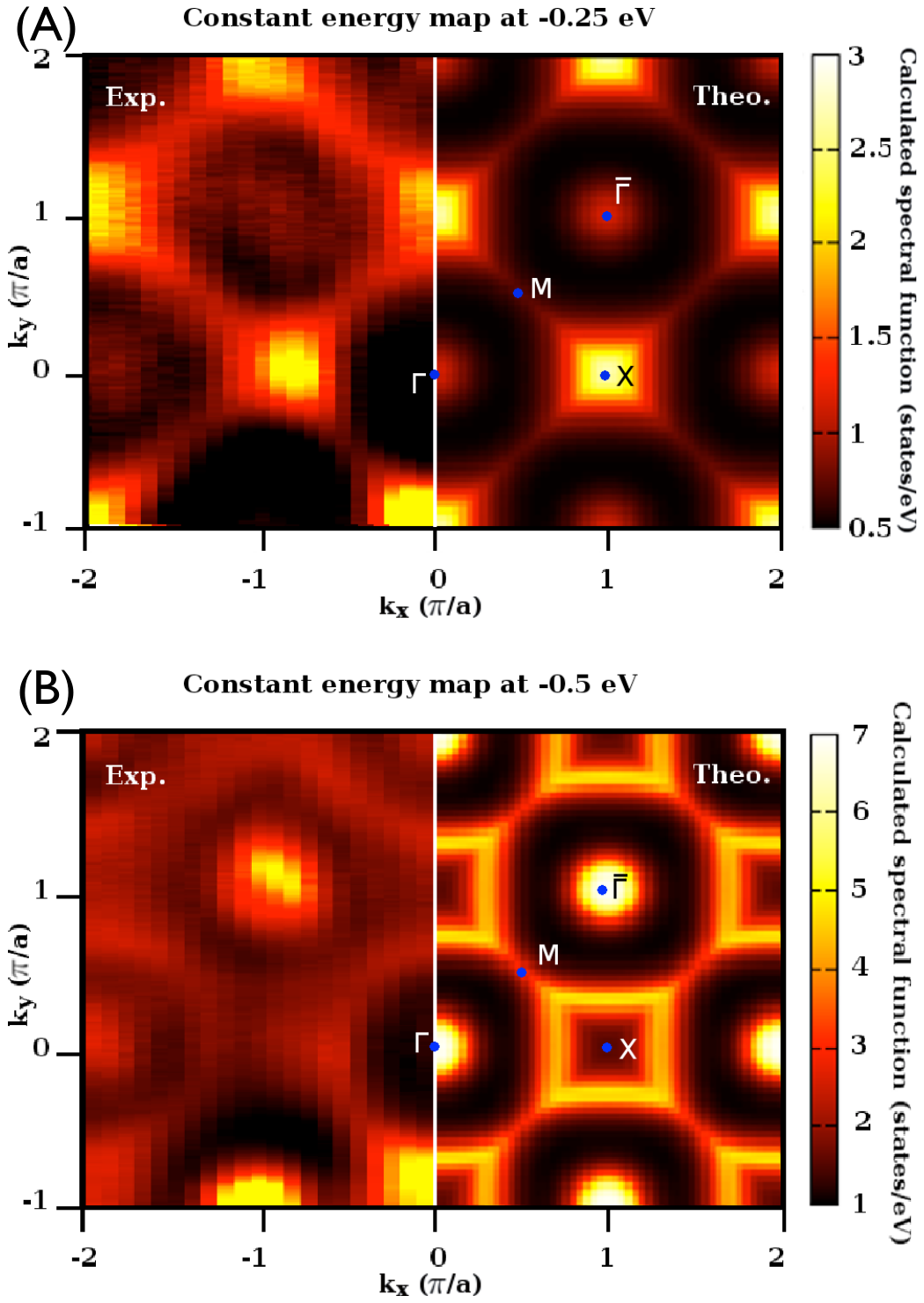}
\caption{\label{ECut}
Constant energy map of the spectral density
of undoped ${\mathrm{Sr}}_{2}{\mathrm{IrO}}_{4}$. 
A: Experimental spectral density (left) and calculated spectral function (right) at $-0.25~\mathrm{eV}$. B: The same quantities at $-0.5~\mathrm{eV}$.
Blue points indicate high-symmetry points of the Brillouin zone.
In the calculations first and second Brillouin zone are the same, in the experiments matrix element effects cause differences.
}
\end{figure}

Our analysis further allows for a refinement of the identification of
the character of the spectral features in the first and second Brillouin zone.
In Fig.~\ref{Akw}C1, the signal comes from the first Brillouin zone and is dominated by the \jeff12 manifold, even though along the $MX$ direction some
spectral weight around $-1.1$~eV originates from the \j32 manifold. 
In Fig.~\ref{Akw}C2, the signal comes from the second Brillouin zone and is dominated by the \j32 bands. 
Our calculations show that even within the \j32 manifold the experimental signal is selective with respect to the $m_j$ quantum number.
The main contribution to the spectral weight seen in the second Brillouin zone comes from the $m_j=1/2$ band whereas the $m_j=3/2$ contributes to the first Brillouin zone.

The left panel of Fig.~\ref{ECut}A shows an ARPES spectrum acquired at
the $X$ point at $-0.25$~eV at room temperature.
According to ARPES measurements of the magnetic phase, the lowest energy excitations disperse up to the $X$ point and never cross the Fermi level. 
This spectrum is qualitatively similar to the one measured below $T_N$ in previous works \cite{KJM+08,WCW+13}, demonstrating that spectral properties are
largely insensitive to the presence or absence of long-range magnetic order.
The lack of a Fermi level crossing excludes the presence of metallic quasiparticles even in the paramagnetic phase, in agreement between theory and experiment.
By comparing the measured spectrum with the calculated spectral function, the overall agreement allows for a clear identification of the peak at $-0.25$~eV with the \jeff12 lower Hubbard band. 
We show in the left panel of Fig.~\ref{ECut}B the photoelectron intensity map collected at $-0.5$~eV. 
The intense blobs observed at $\Gamma$ originate from the top level of the \j32 band whereas the spectral features around $M$ and $X$ arise from the \jeff12 band which form circles centered around $\Gamma$ with a radius of $0.5$~\AA$^{-1}$. 

\subsection{$\mathrm{La}$-doped ${\mathrm{Sr}}_{2}{\mathrm{IrO}}_{4}$ and the putative pseudogap}
\label{ResDoped}

\begin{figure}[b]
\includegraphics[width=0.5\textwidth]{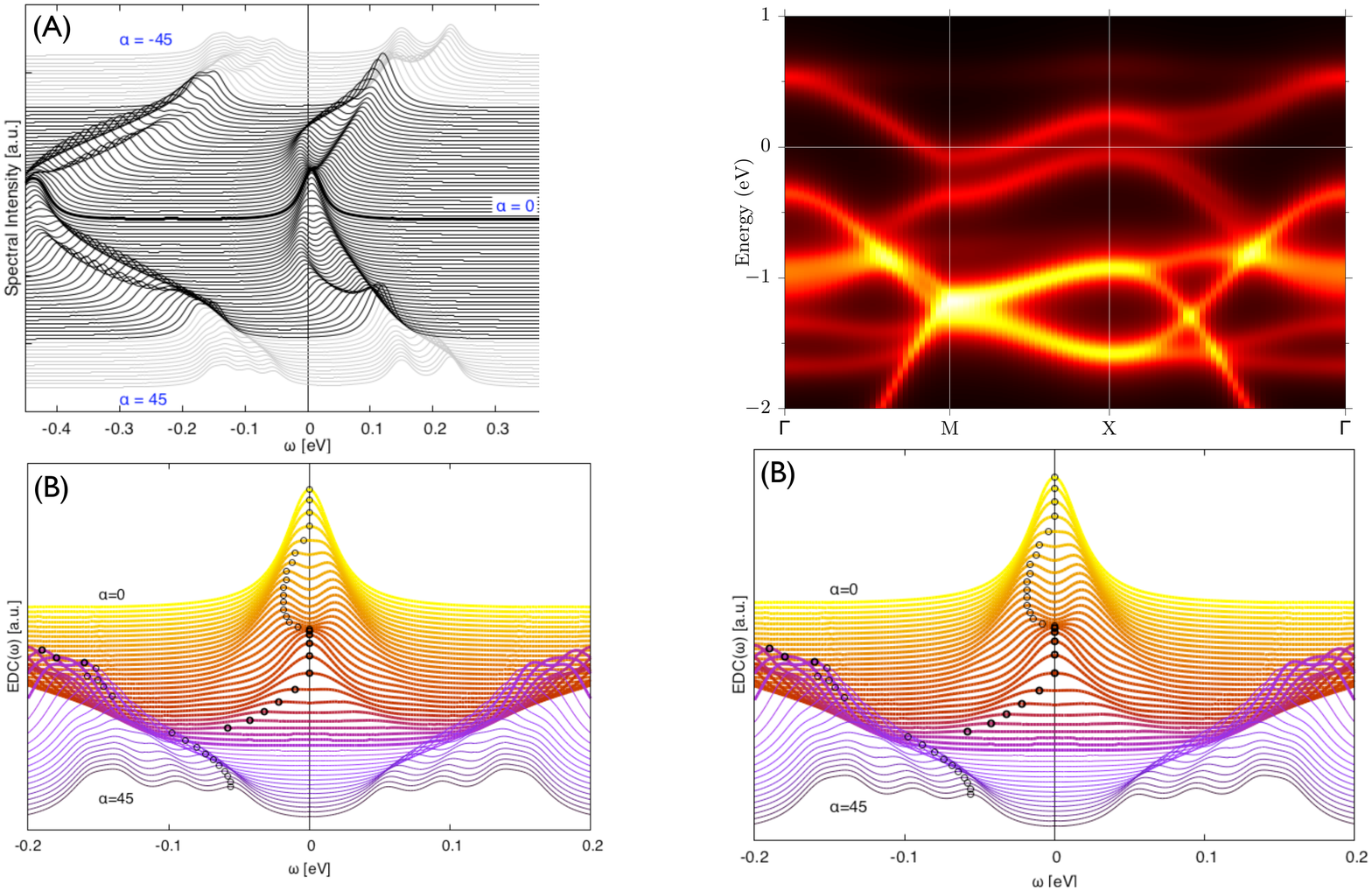}
\caption{\label{Akw_dop}
Calculated momentum-resolved spectral function $A(\mathbf{k},\omega+i\eta)$ of $10\%$ electron-doped Sr$_2$IrO$_4$ along the $\Gamma-M-X-\Gamma$ path. The broadening is $\eta=0.02$~eV.
}
\end{figure}

\begin{figure*}[t]
\includegraphics[width=1.0\textwidth]{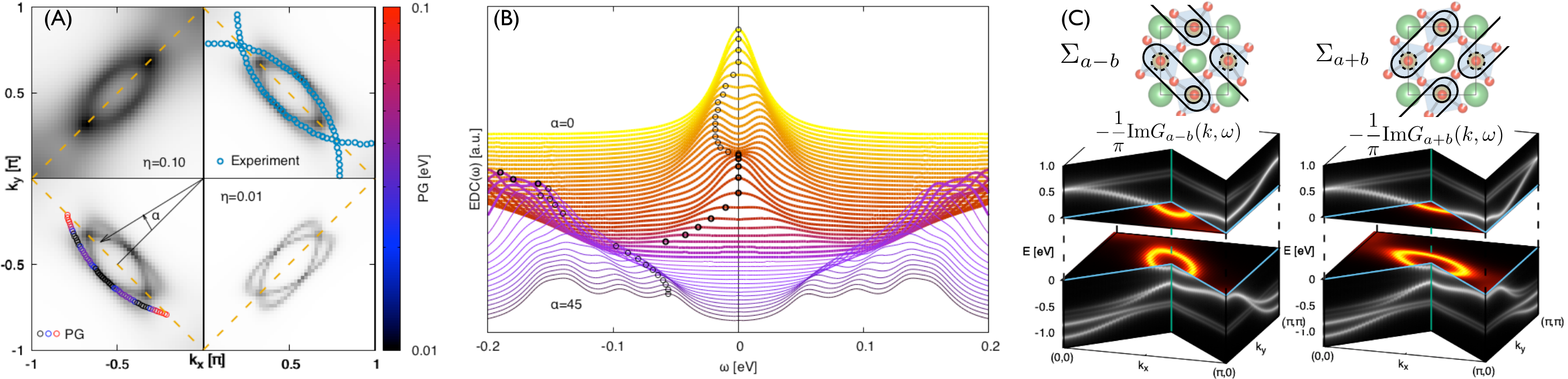}
\caption{\label{Ellipse_EDC}Fermi surface and particle-hole averaged (i.e. ``symmetrized'') spectral functions of $10\%$-electron-doped ${\mathrm{Sr}}_{2}{\mathrm{IrO}}_{4}$. 
A: Fermi surface when combining both dimer configurations. The first Brillouin zone is indicated by a dashed orange line. 
In the first quadrant we overlay experimental data points along which the pseudogap was extracted in Ref.~\cite{dlTMWB+15}, in the third quadrant the pseudogap extracted from (B) along the experimental path. 
B: Symmetrized spectral functions for the semi-circular $\mathbf{k}$-path used in experiment (see (A)). Symmetrized spectral functions for which the pseudogap is shown in (A) are plotted bold and their maxima are indicated by a dot. 
C: Calculated spectral function with equal energy cuts at the Fermi surface for the two cluster orientations $\Sigma_{a-b}$ and $\Sigma_{a+b}$ indicated in the sketches. 
If not indicated in the plots, a broadening of $\eta=0.02$~eV is assumed.}
\end{figure*}

In the following, we investigate spectral properties of the electron-doped system and compare our results to experimental findings on $({\mathrm{Sr}}_{1-x}{\mathrm{La}}_x)_{2}{\mathrm{IrO}}_{4}$ with $x=0.05$ \cite{dlTMWB+15}. In this doping regime the system is a paramagnetic metal down to lowest temperatures.

Figure~\ref{Akw_dop} displays the spectral function $A(\mathbf{k},\omega)$ along the $\Gamma-M-X-\Gamma$ path of the Brillouin zone resulting from our calculations for an electron doping of $10\%$. 
The features of the upper Hubbard band (UHB) of the undoped compound can be continuously connected to features of the quasiparticle states at the Fermi level, which are visible in Fig.~\ref{Akw_dop}.
In our dimer picture, this reminiscent feature of the UHB arises as bonding and antibonding states of the cluster, which is why we will refer in the following to the lower (upper) branch of the \jeff12 band as (anti)bonding band \footnotemark[\value{footnote}].

The antibonding band (ABB) crosses the Fermi level and causes a Fermi pocket around the $M$ point.
The positions of the band maxima at the high symmetry points have changed slightly as compared to the undoped case.
At the $M$ point the bonding band (BB) of the \jeff12 appears around $-0.37~\mathrm{eV}$ and at the $X$ point it is in close proximity to the Fermi energy.
There, depending on resolution 
finite weight of the tail of the broad peak can be picked up at the Fermi energy.
At the $\Gamma$ point the signal of the \j32 manifold can be seen at $-0.4$~eV.

Overall, the spectral function displays excellent agreement with available ARPES data \cite{BMP+15,dlTMWB+15}.
In agreement with Ref. \cite{dlTMWB+15} we find a nearly linear dispersion close to the Fermi level, which can be traced back to the ABB of the \jeff12 band.
Compared to \cite{dlTMWB+15} our spectrum differs slightly at the $M$ point in that in \cite{dlTMWB+15} the dispersion close to the Fermi energy 
was interpreted to extrapolate to a Dirac point around $-0.1$~eV and a quasilinear continuation for even lower energies, possibly with a small gap at the high symmetry line.
Here, we see a distinct gap between bonding and antibonding \jeff12 band at the $M$ point and can clearly distinguish between states of the ABB forming the electron pocket and those of the BB.
Although the lower part of the experimentally measured dispersion matches our calculated spectrum, the states right below the putative Dirac point are not found here.
Our calculation rather suggests a scenario similar to the one of Ref. \cite{BMP+15}, where a similar quasilinear dispersion was found for a doping of $x=0.04$.
There, the vertical dispersion was discussed as stemming from spectral weight of the large tail of the \jeff12, which is picked up by the momentum distribution curve \cite{BMP+15}.


We now turn to the Fermi surface.
The ABB crosses the Fermi energy close to the $M$ point resulting in a lens-shaped electron pocket, 
which is centered around the $M$ point and elongated along the Brillouin zone boundary (see Fig.~\ref{Ellipse_EDC}A).
Our theoretical findings agree well with recent ARPES measurements on the electron-doped system $({\mathrm{Sr}}_{1-x}{\mathrm{La}}_x)_{2}{\mathrm{IrO}}_{4}$, both considering the shape and the size of the pockets \cite{dlTMWB+15}.

The ARPES measurements also revealed unusual metallic behavior \cite{BMP+15,dlTMWB+15} with an antinodal pseudogap at a $\mathrm{La}$ concentration of $x=0.05$ \cite{dlTMWB+15}. 
Figure~\ref{Ellipse_EDC}B shows
particle-hole averaged spectral functions (``symmetrized spectra'') as they
are often plotted from experimental data. We move
along a semicircular $k$-path, which is a natural extension of the Fermi
surface (see Fig.~\ref{Ellipse_EDC}A).
At small angles $4^{\circ}\leq\alpha$ the symmetrized spectral function shows a pseudogap, which closes again at $\alpha\approx15^{\circ}$.
For angles $21^{\circ}\leq\alpha\leq26^{\circ}$ an additional feature emerges at sufficiently large temperatures, which could be misinterpreted as a pseudogap, too.
However, since the hole-part of the spectrum is still close to the Fermi energy, this feature can be traced back to contributions of hole-parts of the spectrum due to large enough temperatures, which then resemble a pseudogap in the symmetrized spectral function.
Finally, for angles $\alpha\lesssim 45^{\circ}$, i.e. close to the antinodal point, spectral features closely resemble the pseudogap features emerging at smaller angles close to the tips of the electron pockets.
In experiment, based on symmetrized energy distribution curves, the regions outside the pockets were identified as showing a pseudogap, which reaches up to the antinodal point \cite{dlTMWB+15}.
Our calculations reveal that the antinodal regions do not show the same pseudogap, which stems from the upper Hubbard band that forms the pockets, but rather relicts of the lower \jeff12 Hubbard band (Fig.~\ref{Akw_dop}). 
To better understand the origin of pseudogap-like features in the symmetrized spectral function, we focus in Fig.~\ref{Ellipse_EDC}A on Fermi surface cuts of the spectral function. 
Within OC-DMFT we have access to the precise position of the quasiparticle excitations. 
However, in plots of the spectral function $A(\mathbf{k},\omega)$ we add a finite broadening $\eta$ to mimic the experimental $\mathbf{k}$ resolution.
The top left panel shows the spectral density at the Fermi level with a realistic broadening $\eta=0.10$~eV. 
It shows the lens-shaped Fermi surface pocket, which is in good agreement with ARPES spectra \cite{dlTMWB+15} (top right panel of Fig.~\ref{Ellipse_EDC}A).
Plotting the extracted pseudogap from the symmetrized spectral function in Fig.~\ref{Ellipse_EDC}B along the semicircular $\mathbf{k}$-path resembles the pseudogap found in ARPES (bottom left panel).

In our calculation, we can trace back the origin of the pseudogap features by choosing an artificially reduced resolution 
(see bottom right panel of Fig.~\ref{Ellipse_EDC}A), which reveals the fine structure of the Fermi pocket.
What seemed to be a sole pocket is actually composed of two separate lens-shaped pockets, which are tilted off-axis in opposite directions by roughly $15^{\circ}$ each.
Given the experimental resolution it is clear that ARPES cannot resolve them separately.
The dimer cluster used in our OC-DMFT calculation is the minimal setup to account for the two different IrO${}_{6}$ orientations of adjacent Ir sites and leads to two configurations (related by a 90 degree rotation), which are schematically shown in Fig.~\ref{Ellipse_EDC}C. 
It is the breaking of orientation symmetry within each configuration, which gives rise to the stretched, tilted Fermi pocket at the Fermi energy as shown in the Fermi surface cuts of the spectral function in Fig.~\ref{Ellipse_EDC}C. 

Since both orientations alternate between the different $\mathrm{Ir}$ layers along the $c$-direction, Figure~\ref{Ellipse_EDC}A shows the Fermi surface of the orientation average of both configurations shown in Fig.~\ref{Ellipse_EDC}C.
A realistic broadening, however, renders it impossible to identify two separate pockets, but rather suggests an interpretation as a single pocket structure (see Fig.~\ref{Ellipse_EDC}A for $\eta=0.10$~eV).
As a consequence, the path along which one extracts the energy distribution curves of this putative single pocket resembles the circular shape of the DFT result that was used in Ref. \cite{dlTMWB+15} and does not coincide with the Fermi surface of the (two-pocket) Fermi surface visible at higher resolution.

Since the putative sole pocket visible at larger broadening is a superposition of the two canted pockets, its tip corresponds to the region between the two pockets. 
When measuring there, the emerging symmetrized spectral function shown in Fig.~\ref{Ellipse_EDC}B features a pseudogap. 
In this sense, the appearance of the pseudogap close to the pocket's tip is a direct consequence of the non-local fluctuations.

\section{Summary and Outlook}
\label{SumOut}
In conclusion, we have performed {\it ab initio} calculations within a novel
oriented cluster dynamical mean field theory scheme, which includes non-local
quantum fluctuations based on an oriented dimer as a reference system for a dynamical mean field theory scheme. We find excellent agreement between the resulting spectral functions for both, pure and electron-doped \sriro, with experimental photoemission data, emphasizing the role of non-local quantum correlations in these 5d compounds. 
These findings provide new evidence for the electronic analogies between iridates and cuprates, making the absence of superconductivity an even more intriguing feature of the iridate compounds.
In particular, the single-orbital nature of a compound, the emergence of a metallic state from a doped insulating parent compound, and the pseudogap-like features discussed in \sriro\ do not 
seem to be proxies for superconductivity.


\acknowledgments

This work was supported by a Consolidator Grant of the European Research
Council (Project CorrelMat-617196), the French Agence Nationale de la
Recherche under project SOCRATE and IDRIS/GENCI Orsay
(Project No. t2017091393).


\bibliography{paper.bib}

\end{document}